\documentclass{book}
\usepackage{krantz_single}%
\usepackage{amsfonts} 
\usepackage{graphicx}

\input epsf		

\begin{document}

\pagestyle{myheadings}
\markboth{Quantum phase transitions in spin-boson systems: dissipation and light phenomena}{Table of Contents}
\tableofcontents
\mainmatter
\markboth{Quantum phase transitions in spin-boson systems}{Quantum phase transitions in spin-boson systems}
\chapter{Quantum phase transitions in spin-boson systems: dissipation and light phenomena}

\vskip -0.7cm
{\bf \Large Karyn Le Hur}
\\
{\it  Departments of Physics and Applied Physics, Yale University, New Haven, Connecticut, 06520, USA}
\\
\\
\\
Spin-boson models are essentially useful in the understanding of quantum optics, nuclear physics, quantum dissipation, and quantum computation. We discuss quantum phase transitions in various spin-boson Hamiltonians, compare, and contrast them. We summarize the theoretical concepts and results, open questions and implementations of those ideas in cold atomic and QED cavity systems will also be addressed. The chapter is organized as follows.

First, a large collection of harmonic oscillators (bosons) can simulate {\it dissipation} in quantum mechanics. Through a two-level system coupled to a bath of bosons, we investigate in detail the concept of  ``dissipation-driven'' quantum phase transition. Another section will be devoted to the effect of dissipation (the bath of bosons) on the critical exponents associated with a well-known phase transition such as the disordered-ordered transition in the Ising model. Second, a spin-boson model can also describe the {\it light}-atom interaction. In particular, the Dicke model describing an ensemble of two-state atoms interacting with a single quantized mode of the electromagnetic field is well-known to exhibit a zero-temperature phase transition at a critical value of the dipole coupling strength. Finally, we theoretically study the superfluid-Mott transition of polaritons in the Jaynes-Cummings lattice system which consists of an array of coupled optical cavities each containing a two-level atom.

\section{Dissipative Transitions for the two-state system}

A dissipative two-state system generally refers to a 
 two-level system coupled to a bath of harmonic oscillators (large collection of bosons) \cite{Leggett}:
\begin{equation}
H = -\frac{\Delta}{2}\sigma_x  + \frac{h}{2} \sigma_z + \frac{1}{2}\sigma_z\sum_{i} c_{i} x_{i} + H_{osc}
\end{equation}
with,
\begin{equation}
H_{osc} = \sum_{i} \left(\frac{p_{i}^2}{2m_{i}} + \frac{1}{2} m_{i} \omega_{i}^2 x_{i}^2\right);
\end{equation}
here, $\sigma_x$ and $\sigma_z$ are Pauli matrices, $\Delta$ is the tunneling amplitude between the
states with $\sigma_z=\pm 1$, and $h$ the bias (detuning).\footnote{Hereafter, the Planck constant  will be set to $\hbar=1$ to avoid confusion with the bias $h$.} Moreover, $x_i$, $p_i$, $m_i$, and $\omega_i$ are the coordinate, momentum, mass, and frequency of the $i$th harmonic oscillator. Here, $c_{i}$ denotes the strength of the coupling with the $i$th oscillator. Information
about the bath is encapsulated in the spectral function \cite{Leggett} $J(\omega) = \frac{\pi}{2} \sum_{i} \frac{c_{i}^2}{m_{i}\omega_{i}}\delta(\omega-\omega_{i})$. This spin-boson model is a variant of the Caldeira-Leggett model \cite{caldeira} where the quantum system
is a spin \cite{Leggett,Blume}. 

\subsection{Ohmic case}

In the case of ohmic dissipation, $J(\omega)=\eta \omega e^{-\omega/\omega_c}$. It is then convenient to introduce the dimensionless dissipation (friction) coefficient $\alpha$ such that
$\eta=2\pi\alpha$. In fact, the emergence of a quantum phase transition can be understood from a perturbation theory in $\Delta/h$ where $h>0$. More precisely, the order parameter obeys $\langle \sigma_z\rangle +1 \sim {\cal O(A)}$ where ${\cal A} = \frac{\Delta^2}{h^2}\left(h/\omega_c\right)^{2\alpha}$ \cite{KLH1}. 
For $\alpha>1$, one observes that the spin tends to be trapped meaning that $\langle \sigma_z\rangle\sim -1$ for $h\ll \omega_c$. In contrast, for $\alpha<1$, ${\cal A}$ increases for smaller $h$ and eventually reaches its maximum value ${\cal A}\sim 1$. This indeed 
suggests the existence of a quantum phase transition at 
$\alpha_c\sim 1$. These impurity systems generally display both a classical (trapped) and quantum (untrapped) phase for the spin \cite{Vojta}.

To better understand the nature of the dissipation-induced transition, one can integrate out the dissipative bath. This leads to an effective action which is reminiscent of the classical spin chains with long-range
correlations \cite{Ising}:
\begin{equation}
\label{Ising}
{\cal S}_{int} = -\int_0^{\beta} d\tau \int_0^{\tau} d\tau' \sigma_z(\tau) {\cal G}(\tau-\tau') \sigma_z(\tau'),
\end{equation}
with ${\cal G}(\tau)\propto 1/\tau^2$ at long (imaginary) times $\omega_c^{-1}\ll \tau\ll \beta=1/k_B T$. It is relevant to observe that Anderson, Yuval, and Hamman \cite{anderson}
found an equivalent Ising model when studying the Kondo problem:\footnote{We obtain the identity $\langle S_z\rangle=\langle \sigma_z\rangle/2$. However, a similar relationship does not hold between $\langle S_x\rangle$
and $\langle \sigma_x\rangle$. The electron operator $c_{k\uparrow}$ must not be confused with
the coupling constant $c_i$ (or $c_k$) in the spin-boson model which will prominently appear in the following.}
\begin{eqnarray}
H_{K}&=&H_{\mbox{\scriptsize kin}}+\frac{J_{\perp}}{2} \sum_{kk^{\prime}} \left( c_{k\uparrow}^{\dagger}c_{k^{\prime}\downarrow} S^- + c_{k\downarrow}^{\dagger}c_{k^{\prime}\uparrow} S^+ \right) \nonumber \\
& & +\frac{J_z}{2}\sum_{kk^{\prime}} \left( c_{k\uparrow}^{\dagger}c_{k^{\prime}\uparrow}-c_{k\downarrow}^{\dagger}c_{k^{\prime}\downarrow} \right) S_z+hS_z,
\end{eqnarray}
where $H_{\mbox{\scriptsize kin}}$ represents the kinetic energy of the electrons. 
This is maybe not so surprising: conduction electrons represent a dissipative bath and as a result tunneling (spin flip) events are not independent. Instead, they feature long-range interactions in time and the equivalent Ising chain acquires
long-range interactions. The trapped-untrapped transition in the spin-boson model is in fact equivalent to the ferromagnetic-antiferromagnetic transition in the anisotropic Kondo model \cite{chakravarty,bray}. The equivalence between the anisotropic Kondo model and the spin-boson model with ohmic damping
can be formulated rigorously through bosonization \cite{Guinea}. The untrapped region (for the spin) corresponds to the antiferromagnetic Kondo model $J_z>0$, while the trapped region corresponds to the ferromagnetic Kondo model $J_z<0$ where the spin is fatally frozen in time. One gets the precise correspondence $(\rho J_{\perp}) \longrightarrow \Delta/\omega_c$ and $(1+2\delta/\pi)^2 \longrightarrow \alpha$,
where $\rho$ is the conduction electron density of states and the parameter $\delta$  is related to the
phase shift caused by the $J_{z}$ Kondo term and is given by $\delta = \tan^{-1}(-\pi \rho J_z/4)$ \cite{Guinea}. Additionally, in the untrapped phase, the effective Kondo energy scale obeys \cite{Bohr}:\footnote{We have introduced a high-energy cutoff $D$ for the conduction electrons which is of the order of the Fermi energy; the relation between $\omega_c$ and $D$ is given, {\it e.g.}, in Refs. \cite{KLH1,Cedraschi}.} 
\begin{equation}
T_K=\Delta (\Delta/D)^{\alpha/(1-\alpha)},
\end{equation}
 for values of $\alpha$ not too close to the transition and close to the transition $T_K$ assumes the exponential form of the isotropic, antiferromagnetic Kondo model; $\ln T_K \propto 1/(\alpha_c-\alpha)$.
The critical line separating the trapped and untrapped phase in the spin-boson model corresponds to $\alpha_c = 1 +{\cal O}(\Delta/\omega_c)$.

Next, we discuss recent developments on spin observables and dynamics.

\subsection{Exact Results}

As shown by Thouless, for the $1/\tau^2$ Ising chain the magnetization $\langle \sigma_z\rangle$ is not
continuous at the transition \cite{Thouless}. This is also consistent with the fact that in the ohmic case, the phase transition is described by Renormalization Group equations similar to those in the XY model in two dimensions. Following Anderson and Yuval, the order parameter $\langle \sigma_z\rangle$
jumps by a {\it non-universal} amount $\sqrt{1/\alpha_c}$ along the quantum critical line, $\alpha_c = 1 +{\cal O}(\Delta/\omega_c)$ \cite{Anderson2}. 

In the untrapped phase $(\alpha<\alpha_c)$, one can apply the Bethe Ansatz approach to compute observables exactly \cite{KLH1,Cedraschi,KLH2}. For $h\ll T_K$, we obtain \cite{KLH1,KLH2}:
\begin{equation}
\lim_{h \ll T_K} \langle \sigma_z \rangle = -\frac{2e^{\frac{b}{2(1-\alpha)}}}{\sqrt{\pi}}  \frac{\Gamma[1+1/(2-2\alpha)]}{\Gamma[1+\alpha/(2-2\alpha)]} \left( \frac{h}{T_K} \right),
\end{equation}
where $b=\alpha \ln \alpha + (1-\alpha) \ln (1-\alpha)$. Note that $\langle \sigma_z \rangle \propto h/T_K$ at small $h$, in keeping with the Kondo Fermi liquid ground state \cite{Leggett}. The local
susceptibility of the spin converges to $1/\Delta$ for $\alpha\rightarrow 0$\footnote{For $\alpha\rightarrow 0$, $\Gamma[1]=1$, $\Gamma[3/2]=\sqrt{\pi}/2$, $\exp(b/(2(1-\alpha))=1$, and $T_K=\Delta$.}  in accordance with the
two-level description and diverges in the vicinity of the phase transition due to the exponential suppression of $T_K$. Note that the longitudinal spin magnetization (the spin order parameter) $\langle \sigma_z\rangle$ only depends on the ``fixed point'' properties, {\it i.e}, this is a universal function of $h/T_K$ in the untrapped phase. Finally, in the trapped phase, one predicts $\langle \sigma_z\rangle \approx -1 +{\cal O}((\Delta/\omega_c)^2)$ \cite{Angela}.

The leading behavior of $\langle \sigma_x \rangle$ in the untrapped phase is \cite{KLH1,Cedraschi,KLH2}:
\begin{equation}
 \lim_{h \ll T_K} \langle \sigma_x\rangle = \frac{1}{2\alpha-1} \frac{\Delta}{\omega_c}+M(\alpha) \frac{T_K}{\Delta},
 \label{sigxh0}
 \end{equation}
 with 
\begin{equation}
M(\alpha)  =  \frac{e^{-b/(2-2\alpha)}}{\sqrt{\pi}(1-\alpha)} \frac{\Gamma[1-1/(2-2\alpha)]}{\Gamma[1-\alpha/(2-2\alpha)]}.
\end{equation}

\begin{figure}
\resizebox{\hsize}{!}{
\includegraphics{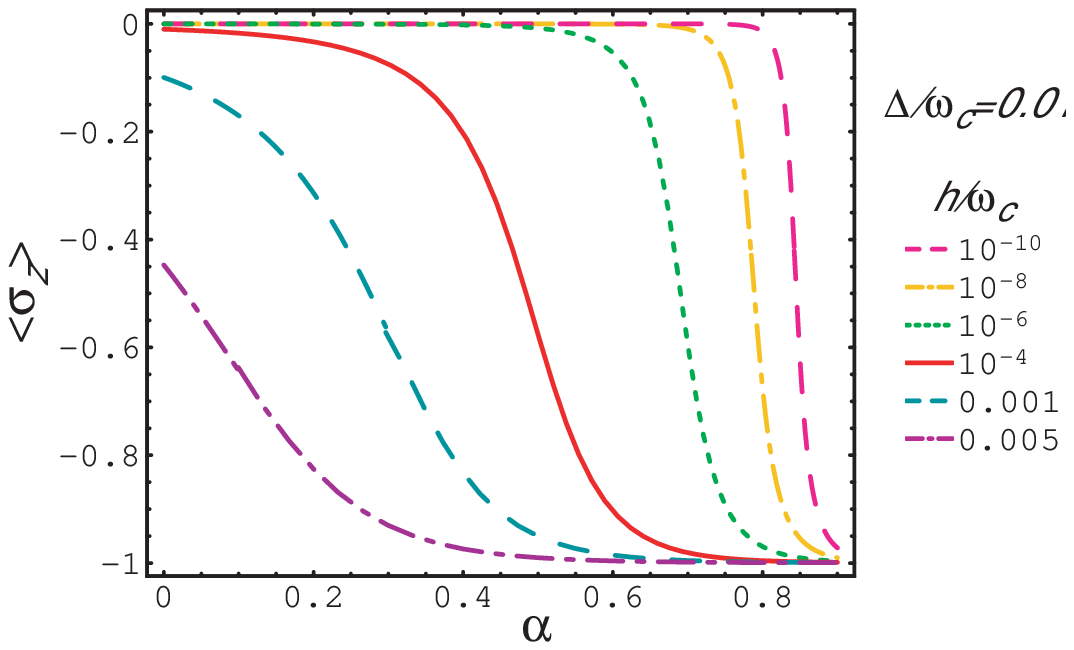}
\includegraphics{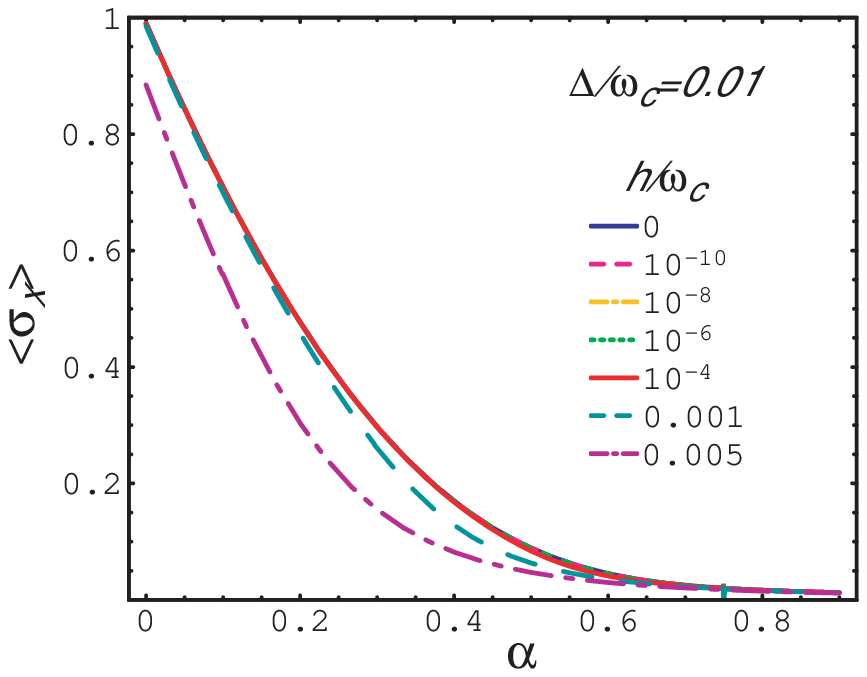}
}
\caption{Longitudinal and Transverse spin magnetizations versus $\alpha$ from Bethe Ansatz.}
\label{fig:un}
\end{figure}

As $\alpha \to 0$, $T_K \to \Delta$ and $M(0)=1$, so we check that $\langle \sigma_x \rangle\rightarrow 1$. As we turn on the coupling to the environment, we introduce some uncertainty in the spin direction and $\langle \sigma_x \rangle$ decreases. Note that $\langle \sigma_x\rangle$ does not only depend on the fixed point properties; more precisely, in the untrapped (and highly non-perturbative) regime, $\langle \sigma_x\rangle$ still contains a perturbative part in $\Delta/\omega_c$ stemming from
the trapped region! For $\alpha < 1/2$, the monotonic decrease of $T_K/\Delta$ dominates.  In contrast, for $\alpha>1/2$, the first term in Eq.~(\ref{sigxh0}) dominates:
\begin{equation}
\langle \sigma_x\rangle_{\alpha>1/2,h\rightarrow 0} = \frac{1}{2\alpha-1}\frac{\Delta}{\omega_c}.
\end{equation}
This result can also be recovered using a perturbation theory in $\Delta/\omega_c$ \cite{KLH1}. This emphasizes that the observable $\langle \sigma_x\rangle$ is {\it continuous} and {\it small} at the quantum phase transition. This is also consistent with the work by Anderson and Yuval which predicts 
$\langle \sigma_x\rangle \sim \Delta/\omega_c$ exactly at the quantum phase transition \cite{Anderson2}. 

Finally, we can also check that the spin component $\langle \sigma_x\rangle$ evolves continuously close to $\alpha=1/2$. 
In the limit $\alpha \to 1/2$, one can take $M(\alpha) = (4/\pi) \Gamma(1-2\alpha) \to 4/(\pi (1-2\alpha))$ and use the identity $D(\alpha=1/2)=4\omega_c/\pi$ \cite{KLH1,Cedraschi} to find 
\begin{equation}
\langle \sigma_x \rangle \to -(4/\pi) \sqrt{T_K/D} \ln (T_K/D), 
\end{equation}
in agreement with the ``non-interacting'' resonant level description valid at the specific point 
$\alpha=1/2$ \cite{KLH1}. Exact results for the spin observables obtained using Bethe Ansatz are summarized in
Fig. \ref{fig:un}. Usually,  a strong reduction
of the off-diagonal diagonal elements of the spin reduced density matrix traduces quantum decoherence. Using the results above, we observe that quantum decoherence is prominent
for $\alpha\geq 1/2$ (where $\langle \sigma_x\rangle$ becomes tiny and the entanglement between the spin and the bath becomes almost maximal \cite{KLH1,KLH2}). 

\subsection{Spin dynamics and Entanglement}

Another useful quantity in the context of macroscopic quantum coherence is the occupation probability
$P(t)=\langle \sigma_z(t)\rangle$, where the system is subject to the non-equilibrium initial preparation $\sigma_z(t=0)=+1$ and the initial density matrix is in a factorized form \cite{Leggett}. At time $t=0$, the dynamics starts out. 

To study the spin dynamics it is convenient to perform a polaronic transformation $U=\exp(-i \sigma_z \Omega/2)$ where $\Omega=\sum_i (c_i/m_i\omega_i^2)p_i$, such that the transformed Hamiltonian
 $H'=U^{-1}HU$ takes the precise form  \cite{Leggett}
\begin{equation}
H' = -\frac{1}{2}\Delta\left(\sigma_+ e^{-i\Omega} +\sigma_-e^{i\Omega}\right)+ \sum_{i} \left(\frac{p_{i}^2}{2m_{i}} + \frac{1}{2} m_{i} \omega_{i}^2 x_{i}^2\right).
\end{equation}
In the Heisenberg picture, the equations of motion for $\sigma_{\pm}(t)$ are easily obtained. Integrating
and substituting them into the equation of motion for the transverse polarization $\sigma_x(t)$, then one gets the exact
formula:
\begin{equation}
\label{spind}
\dot{\sigma}_z(t)=-\frac{1}{2}\Delta^2 \int_{-\infty}^t \left( e^{-i\Omega(t)}e^{i\Omega(t')} \sigma_z(t')
+\sigma_z(t')e^{-i\Omega(t')}e^{i\Omega(t)}\right)dt'.
\end{equation}
On the other hand, to solve this equation, one usually uses approximations \cite{Dekker}. The first approximation generally consists to insert the free bath dynamics when computing the commutator:
\begin{equation}
[\Omega(t),\Omega(t')] = i \sum_j \left(\frac{c_j^2}{m_j\omega_j^3}\right)\sin(\omega_j(t-t')).
\end{equation}
The next step is to average (\ref{spind}) with respect to the bath and to decouple the environmental exponentials from the spin. Using that:
\begin{equation}
\langle \Omega(t)\Omega(t') + \Omega(t')\Omega(t)\rangle = \sum_j \frac{c_j^2}{m_j\omega_j^3}
\coth\left(\frac{1}{2}\beta\omega_j\right)\cos(\omega_j(t-t')),
\end{equation}
this leads to the evolution equation \cite{Dekker}:
\begin{equation}
\dot{P}(t) +\int_{-\infty}^t {\cal F}(t-t')P(t')dt' =0,
\end{equation}
where the function ${\cal F}$ obeys ${\cal F}(t) = \Delta^2 \cos\left(A_1(t)\right) \exp-\left(A_2(t)\right)$, and
\begin{eqnarray}
A_1(t) &=& \frac{1}{\pi}\int_0^{+\infty} \sin(\omega t) \frac{J(\omega)}{\omega^2} d\omega \\ \nonumber
A_2(t) &=& \frac{1}{\pi}\int_0^{+\infty} \left(1-\cos(\omega t)\right) \coth\left(\frac{\beta \omega}{2}\right)
\frac{J(\omega)}{\omega^2} d\omega.
\end{eqnarray}
Through the Laplace transform one obtains ($C$ denotes a Bromwich contour):
\begin{equation}
P(t) = \frac{1}{2\pi i} \int_C d\lambda e^{\lambda t} \frac{1}{\lambda+{\cal F}(\lambda)}.
\end{equation}
At zero temperature and in the scaling limit $\Delta/\omega_c\ll 1$, one finds \cite{Leggett}:
\begin{equation}
{\cal F}(\lambda) = \Delta_e\left(\frac{\Delta_e}{\lambda}\right)^{1-2\alpha},
\end{equation}
where $\Delta_e = \Delta_r \left(\cos(\pi\alpha)\Gamma(1-2\alpha)\right)^{\frac{1}{2(1-\alpha)}}$; 
we have introduced the renormalized transverse field $\Delta_r = \Delta(\Delta/\omega_c)^{\alpha/1-\alpha}$ which is proportional to the Kondo energy scale $T_K$. This expression of $P(t)$ coincides with the formula
of $P(t)$ obtained via the 
Non-Interacting Blip Approximation (NIBA) \cite{Leggett}. 

For $\alpha\rightarrow 0$, one recovers perfect Rabi oscillations $P(t)=\cos(\Delta t)$ whereas for $\alpha=1/2$ one gets a pure {\it relaxation} $P(t) = \exp-(\pi\Delta^2 t/(2\omega_c))$, which is in accordance with the non-interacting resonant level model \cite{Leggett}.
For $0<\alpha<1/2$, the spin displays  coherent oscillations (due to a pair of simple poles) leading to $P_{coh}(t)=a\cos(\zeta t+\phi)\exp(-\gamma t)$ with the quality factor \cite{Leggett}:
\begin{equation}
\label{quality}
\frac{\zeta}{\gamma} = \cot\left(\frac{\pi\alpha}{2(1-\alpha)}\right).
\end{equation}
This quality factor has also been found using conformal field theory \cite{Lesage}.

\begin{figure}
\resizebox{\hsize}{!}{
\includegraphics{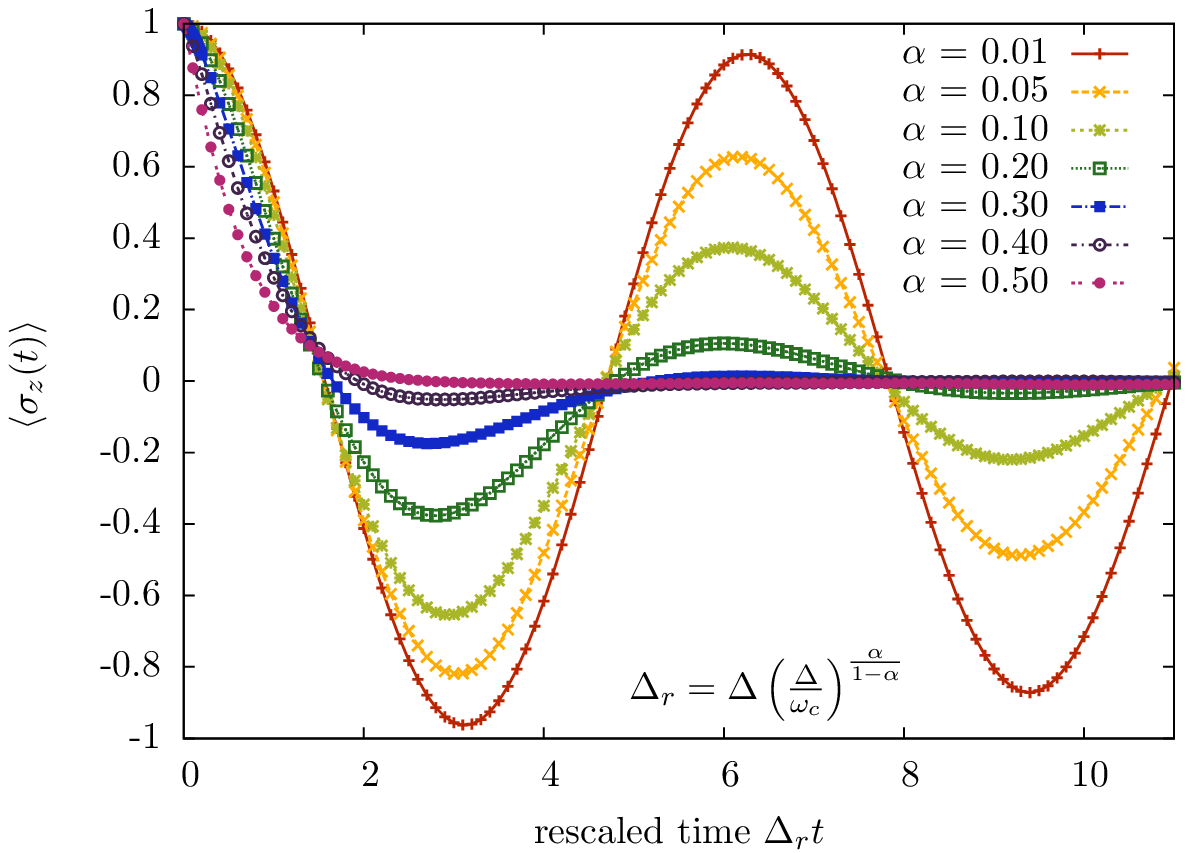}
\includegraphics{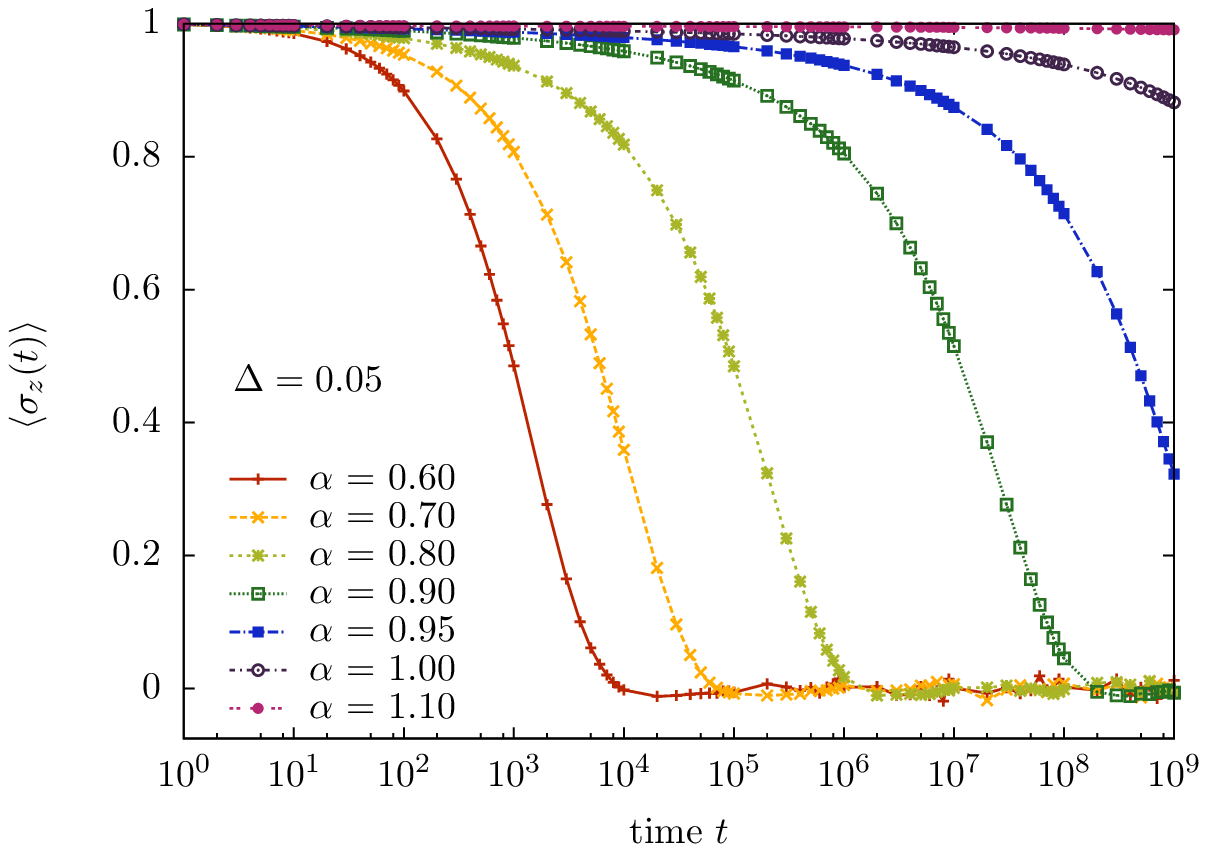}
}
\caption{Spin dynamics $P(t)=\langle \sigma_z(t)\rangle$ obtained from time-dependent NRG.}
\label{fig:deux}
\end{figure}

Recently, we have developed a time-dependent Numerical Renormalization Group (NRG) which allows
us to confirm these results \cite{David}; see Fig. \ref{fig:deux}. In particular, we
have checked the expression given in  (\ref{quality}) for the quality factor. In fact, the NRG represents a powerful theoretical tool to study those spin-boson systems \cite{Bulla,Walter,Anders}. Additionally, we have obtained the same quality factor using an exact analytical extension of the NIBA approach \cite{PeterAdilet}.

It is relevant to note that the coherent-incoherent crossover, corresponding either to the
strong suppression of the off-diagonal elements of the spin reduced density matrix $\langle \sigma_x\rangle \sim \Delta/\omega_c\rightarrow 0$ or to the complete
vanishing of the Rabi quantum oscillations, can be identified to the Toulouse limit $\alpha=1/2$ and not
to the quantum phase transition. It is also interesting to underline the correspondence between the prominence
of spin-bath entanglement and the emergence of quantum decoherence \cite{KLH1,KLH2}. The entanglement entropy $S$ between the spin and the environment displays a plateau at maximal entanglement for $1/2<\alpha<1$. Using the time-dependent NRG \cite{David}, we have also studied the strong coupling regime and in particular the crossover from incoherent decay to localization 
at the quantum phase transition (Fig. \ref{fig:deux}). In particular, for $1/2<\alpha<1$, the spin dynamics
remains purely incoherent. The authors of Ref. \cite{WangD} report a multi-exponential form $P(t)=\sum_{n=1}^{+\infty} a_n e^{-2b_n t}$. In contrast with the results of conformal field theory \cite{Lesage}, the rate  constants appearing in the multi-exponential fit, are not integer multiples of a single rate. Our NRG results rather support the form $P(t)=\exp[(- t T_K/2)^{a_{\alpha}}]$ with the prerequisite
that $a_{1/2}=1$. The exponent $a_{\alpha}$ evolves linearly with $\alpha-0.5$.

\subsection{Sub-ohmic case}
\label{longIsing}

Now, we focus on the sub-ohmic situation $J(\omega)=2\pi\alpha\omega_c^{1-s}\omega^s$ with $0<s<1$ which exhibits a second-order quantum phase transition \cite{Vojta} by analogy to classical spin chains \cite{Ising,Dyson}: in this case ${\cal G}(\tau)\propto 1/\tau^{1+s}$ in Eq. (\ref{Ising}). The second-order quantum phase transition separates a localized (trapped) phase for the spin at small $\Delta$ from a delocalized (untrapped) phase at large $\Delta$.

For a second-order impurity quantum phase transition we can apply the following scaling ansatz for the impurity part of the free energy \cite{Vojta},
\begin{equation}
\label{free}
F_{imp} = T F(|\Delta-\Delta_c|/T^{1/\nu}, h T^{-b}),
\end{equation}
where we have re-introduced the detuning $h$ and $\Delta_c$ is the value of the transverse field at the quantum critical point.  The critical exponent $b$ should not be confused with the parameter $b$ of Sec. 1.1.2. There is no independent dynamical critical exponent for $(0+1)$ dimensional models, formally $z=1$. At zero temperature, the crossover from the quantum critical regime to one or other of the stable regimes, defines an energy scale $h^*$ which vanishes at 
$\Delta_c$, $h^*\propto |\Delta_c-\Delta|^{b\nu}$. 
In a similar way, at finite temperature, we define the energy scale 
$T^*= |\Delta_c-\Delta|^{\nu}$.
It should be noted that the ansatz (\ref{free}) is usually well justified when the fixed point is interacting \cite{Vojta}; for a Gaussian fixed point the scaling function would also depend upon dangerously irrelevant variables. 

Both analytical arguments, based on the equivalence to a O(1) $\phi^4$ theory
and numerical simulations for the one-dimensional long-range Ising model show that the upper-critical
dimension is $d_u=2s$ \cite{Fisher,Blote}. In other words, the transition obeys non-trivial critical behavior for $1/2<s<1$ and the fixed point is interacting. More precisely, hyperscaling relations imply that there are only two independent exponents, {\it e.g.}, $\nu$ and $b$. A Ward identity for the spin-boson model ensures that $b=(1+s)/2$ \cite{KLH1,Philippe,Zarand}. Defining
the exponent $\delta$ as $\langle \sigma_z\rangle (h,\Delta_c) \propto |h|^{1/\delta}$, one also
finds $\delta=(1+s)/(1-s)$. This implies that the local susceptibility $\partial \langle \sigma_z\rangle/\partial h$ at the quantum critical point diverges as $T^{-s}$. Hyperscaling also guarantees an $\omega/T$ scaling, $\chi_{loc}''(\Delta_c,T=0,\omega)\propto |\omega|^{-s}\hbox{sgn}(\omega)$. Notably, this was found to be the exact decay exponent of the critical spin correlations in the long-range Ising model for all $s$ \cite{Fisher}. Additionally, the exponent $\nu$ diverges as
$1/\sqrt{2(1-s)}$ near the ohmic point $s=1$ \cite{Vojta}. For $0<s<1/2$, the transition is mean-field like and critical exponents obey, {\it e.g.}, $\delta=3$ and $\nu=1/s$ \cite{Fisher,Blote}. Hyperscaling is violated. Very recently, using a powerful 
continuous time cluster Monte-Carlo algorithm, Winter {\it et al.} have shown 
that the quantum-to-classical mapping is valid in the case of 
the sub-ohmic spin-boson model \cite{VojtaMC}. On the other hand, note that the presence of a dangerously irrelevant variable for $s<1/2$ impedes the correct extraction of the critical exponents with current versions of the  NRG method.

In Refs. \cite{KLH1,Philippe}, we have studied in detail the entanglement properties in the sub-ohmic spin-boson model. In particular, for those second-order quantum phase transitions, the entanglement entropy exhibits a visible cusp at the quantum critical point and quantum decoherence becomes maximized.

\subsection{Realizations}

The spin-boson model can be realized in noisy charge qubits built of mesoscopic quantum dots or Cooper pair boxes \cite{Schon,Clerk}. The gate voltage controls the detuning $h$ and $\Delta$ corresponds to the tunneling amplitude between the dot and the lead(s) or the Josephson coupling energy of the junction. If the gate voltage source is placed in series with an external resistor, which can be modelled by a long $LC$ transmission line, this may describe the spin-boson model with ohmic dissipation \cite{Cedraschi}. A one-dimensional Luttinger reservoir could also be used \cite{KLH3,Matveev}. The spin-boson model
can also be derived when coupling a quantum dot to a boson and a fermion bath \cite{KLH4}. These nano-systems may also allow to address new important issues such as the non-equilibrium transport properties at a given quantum phase transition \cite{Chung}. The sub-ohmic case $s=1/2$ can be engineered through an $RLC$ transmission line.
Charge measurements could provide the quantity $\langle \sigma_z\rangle$, which represents the occupation of the dot or island. In a ring geometry, the application of a magnetic flux generates a persistent current
which is proportional to $\langle \sigma_x\rangle$ \cite{Cedraschi}. Solid-state two-level systems usually feature a coupling strength much below $\alpha=1/2$. A very promising candidate is the ultracold-atomic quantum dot coupled to a Bose-Einstein Condensate (BEC) \cite{Recati}, which allows an {\it unprecedented} control of the coupling(s) between the qubit and the reservoir. The spin-boson model can also be engineered in trapped ions arranged in Coulomb crystals \cite{Cirac}.

\section{Dissipative spin array}

Now, we investigate the ground state of a spin array coupled to a common (large) collection of harmonic oscillators. We intend to show that when the coupling is longitudinal the system can be mapped onto a dissipative quantum Ising model. To simplify the discussion, we consider a one-dimensional channel model for the bath and allow bosons (phonons, sound waves in a BEC, or photons) to propagate along a single direction with wavevector $k$ and dispersion $\omega_k = v k$. More general results are shown in Ref. \cite{Peter}. Similar to the single spin case, the interaction between the spins and the boson bath reads,
\begin{equation}
H_{Int} = \sum_i \sum_{k} \frac{c_k}{2} e^{i {k x_i}} (a_{k} + a_{-k}^{\dagger}) \sigma_{iz},
\end{equation}
whereas the boson bath Hamiltonian reads $H_B=\sum_k \omega_k a^{\dagger}_k a_k$ (we set $\hbar=1$). In this equation, $x_i$ correspond to the positions of the spin impurities.

\subsection{Boson-mediated magnetic interaction}

First, we show that an exchange interaction between the spins is induced by the bosonic environment; this is analogous to the Ruderman-Kittel-Kasuya-Yosida interaction induced by a fermionic bath. For this purpose, we set the transverse field $\Delta=0$. In fact,
the spin-boson interaction can be exactly eliminated through a polaronic (unitary) transformation 
along the lines of the single spin case \cite{Leggett}. More precisely, we perform a unitary transformation $V=\exp A = \exp
\left(\sum_i\sum_k \frac{A_k}{2} e^{ik x_i}(a_k-a^{\dagger}_{-k})\sigma_{iz}\right)$. The transformed Hamiltonian then takes the general form (where $H=H_{Int}+H_B+\sum_i \frac{h}{2}\sigma_{iz}$):
\begin{equation}
e^{-A} H e^{A} = H + [H,A] +\frac{1}{2}[[H,A],A]+...\ .
\end{equation}                                  
Now, we choose the coefficients $A_k=c_k/w_k$ such that the induced term $[H_B,A]$ exactly cancels
$H_{Int}$. On the other hand, the transformed Hamiltonian also produces an effective interaction between spins:
\begin{equation}
\delta H = -\frac{1}{4}\sum_{i,j}\sum_k \frac{c_k^2}{\omega_k} e^{ik(x_i-x_j)} \sigma_{iz}\sigma_{jz}.
\end{equation}
If one envisions a cold atomic spin array coupled to a BEC reservoir, one can check that the factor $c_k^2/\omega_k$ is
k-independent \cite{Peter}. Furthermore, we identify:
\begin{equation}
\label{sinc}
\sum_k e^{ik(x_i-x_j)} = \frac{L}{\pi \xi_h}\hbox{sinc}((x_i-x_j)/\xi_h);
\end{equation}
we have introduced the function $\hbox{sinc}(x)=\sin(x)/x$ and $\xi_h=v/\omega_c$ ($\omega_c$
represents the ultraviolet cutoff for the sound modes in the BEC and $L$ denotes the length of the BEC). 
Notice that the induced
interaction decays very rapidly for separations larger than the healing length $\xi_h$. In a realistic
cold atom experiment, the distance between (atomic) quantum dots is of comparable size as the healing length. Therefore, we can restrict ourselves to nearest-neighbor spin interactions. It should also be noted that the effective
interaction $-K<0$ is {\it ferromagnetic} and independent of the length $L$ since $c_k^2/\omega_k\propto 1/L$ \cite{Peter}.

\subsection{Solvable dissipative model}

On the other hand, the boson bath is also expected to give rise to long-range correlations in time similar
to the single spin case. In fact, to find those long-range correlations in time, it is judicious to integrate out the
phonon (sound wave) modes using coherent state functional integrals. At a general level, this produces the following extra term in the action of the spin array \cite{Peter}:\footnote{We use the convention $\sigma_k$ instead of $\sigma_{kz}$ to save space.}
\begin{equation}
\delta S = - \frac{c_k^2}{4 \beta \omega_k} \sum_{k; n=-\infty}^{\infty} \int_0^{\beta} d\tau d\tau' \left(\frac{\omega_k^2 - i \omega_k \Omega_n}{\omega_k^2 + \Omega_n^2} \right) e^{i \Omega_n (\tau - \tau')} \sigma_{k}(\tau) \sigma_{k}^*(\tau'),
\end{equation}
where $\beta=(k_B T)^{-1}$, $\sigma_{k}(\tau) =\sum_j \sigma_{jz}(\tau) e^{i k x_j}$, and $\Omega_n=2\pi n/\beta$ are Matsubara frequencies. Interestingly, the coupling to the boson bath provides two {\it distinct} contributions, which can be identified using the decomposition, $\sigma_{k}(\tau) \sigma_{k}^*(\tau') = \frac12 (\sigma_{k}(\tau) \sigma_{k}^*(\tau) + \sigma_{k}(\tau') \sigma_{k}^*(\tau')) - \frac12 (\sigma_{k}(\tau) - \sigma_{k}(\tau')) (\sigma_{k}^*(\tau) - \sigma_{k}^*(\tau'))$ \cite{Leggett}. 

The first term which is local in time is dominated by the Matsubara frequency term $n=0$. This allows us to recover
the ferromagnetic Ising contribution $\delta H$ found above by applying the unitary transformation $V$. 

Now, similar to the case of a single two-level system coupled to a bosonic bath \cite{Leggett},  the second (dissipative) contribution stems from the pole at 
$\omega_{k}=-i\Omega_n$ (or $\omega_{k}=i\Omega_n$). One can check that the main dissipative contribution is for $i=j$. This term produces an on-site long-range correlation in time $\propto 1/\tau^2$.
Note that correlations in space are short-range, as a result of Eq. (\ref{sinc}), whereas the correlations in time are long-range due to dissipation.

Re-introducing the transverse field $\Delta$, then we find that the spin dynamics is dictated by the following action \cite{Peter}:
\begin{eqnarray}
 S_{spin} &=&  - \sum_i \int_0^{\beta} d\tau \left[\frac{\Delta}{2} \sigma_{ix}(\tau) + \frac{h}{2} \sigma_{iz}(\tau) \right] \\ \nonumber
    & - &
    \frac{1}{8} \sum_i \int_0^{\beta}  d\tau d\tau' \alpha(\tau - \tau') \sigma_{iz}(\tau) \sigma_{iz}(\tau')\\
    \nonumber
&-& K  \sum_{i,j} \int_0^{\beta} d\tau \;  \sigma_{iz}(\tau)  \sigma_{jz}(\tau).
\end{eqnarray}
We have properly defined:
\begin{equation}
\alpha(\tau - \tau') = \frac{1}{\pi} \int_0^\infty d\omega J(\omega) e^{- \omega |\tau - \tau'|},
\end{equation}
and the spectral function reads $J(\omega) = \pi \sum_k c_k^2 \delta(\omega - \omega_k)$. In fact, one can establish that $J(\omega)=2\pi\alpha\omega^d$ where $d$ denotes the dimensionality of the BEC. 

Assuming we consider two-state systems close to the degeneracy point $h=0$, then one recognizes
the action of a {\it dissipative quantum Ising model} \cite{Matthias}. 

\subsection{Dissipative $\phi^4$ theory}

For $\Delta=0$ the system is in a ferromagnetic (ordered) phase whereas for large $\Delta\gg K$ the system will be in a paramagnetic (disordered) phase. The nondissipative quantum Ising model ($\alpha=0$) exhibits a second order phase transition around $K\sim\Delta$, separating the paramagnetic and ferromagnetic phase, and can be mapped exactly onto the classical Ising model in $d_{eff} = d+z = d+1$ dimensions; $z=1$ is the dynamic critical exponent that describes the relative dimensions of (imaginary) time and space. The classical model itself can be described by a $\phi^4$ theory. In particular, the quantum Ising chain lies in the universality class of the two-dimensional classical model
 \cite{Onsager}. 
 
When $\alpha>0$, the critical behaviour is most profoundly changed in the ohmic case ($d=1$). 
 First, as a reminiscence of the single two-state system, the bath will renormalize the transverse field as
 $\Delta_r=\Delta \left(\Delta/\omega_c\right)^{\alpha/(1 - \alpha)}$ \cite{Leggett}, such that the transition occurs at a smaller $K\sim\Delta_r$. Second, for $d=1$, the dissipation will generate a term $|\omega|\phi(q,i\omega)^2$ in the effective $\phi^4$
 theory which will affect the critical exponents and change the universality class of the phase transition. 
 
\subsection{Critical exponents}

Here, we summarize the main results for the critical exponents.  In fact, the latter have been shown to be independent of the value of the parameter $\alpha$ \cite{Matthias,Matthias2} and they have been thoroughly derived through a dissipative $\phi^4$ theory in $d= (2 -\varepsilon)$ dimension ($\varepsilon=1$) \cite{Matthias2,Georges} and Monte Carlo simulations \cite{Matthias}. 

\begin{table}[t]
\tabletitle{Critical exponents of (dissipative) quantum Ising models.}
  \begin{tabular}{|c|c|c|c|c|c|c|}
 Critical Exponents & $\beta$ & $\gamma$ &$\delta$ & $\nu$ & $z$ \\
  \hline
  Dissipative Quantum Chain & $0.319$ & $1.276$ & $\sim 5$ & $0.638$ & $1.98$ \\ \hline
  $d=2$ Quantum Ising model & $0.325$ & $1.241$ & $4.82$ & $0.630 $ & $1$ \\ \hline
  Non-dissipative Quantum Chain & $1/8$ & $7/4$ & $15$ & $1$ & $1$ \\ 
    \end{tabular}
\label{tab:1}
\end{table} 

For $d=1$, due to the dissipative term $|\omega|\phi(q,i\omega)^2$ in the effective O(1) 
$\phi^4$ theory, 
the dynamic critical exponent is equal to $z\approx 2$, whereas $z=1$ for $d=(2,3)$.  
Therefore, the one-dimensional dissipative and two-dimensional quantum Ising models should behave similarly, since they have the same dimension $d_{eff} = d + z \approx 3$. For $d=1$,
the $\phi^4$ theory (up to second order in $\varepsilon$) predicts $z\approx 1.98$ \cite{Georges} 
which is in accordance
with the Monte Carlo simulations \cite{Matthias,Matthias2}. A summary of some critical exponents
for the one-dimensional dissipative (quantum) case and the two-dimensional quantum Ising model is presented in Table \ref{tab:1}. The exponent $\delta$ is defined in Sec. \ref{longIsing}. Additionally,
$\sigma_{iz} \propto \left|\Delta-\Delta_c\right|^{\beta}$ and $\chi = d\sigma_{iz}/d h \propto 
|\Delta-\Delta_c|^{-\gamma}$.  The correlation length exponent obeys $\nu\approx 0.638$ \cite{Matthias,Matthias2}. In addition, from scaling laws, 
$\gamma=z\nu\approx 1.276$ \cite{Matthias, Matthias2, Georges} and $\beta=\nu/2\approx 0.319$.
One must also satisfy $z=(\delta-1)/2$. 

\subsection{Realizations}

In Ref. \cite{Peter}, we have shown that a spin-boson mixture of cold atoms can be used to engineer the quantum Ising model in a dissipative bath. We emphasize that this setup embodies the first tunable realization of the quantum Ising model in a dissipative bath, and that several critical exponents can be measured using standard imaging techniques. On the other hand, the effect of a (nuclear spin) bath on the quantum phase transition of an Ising ferromagnet in a transverse field has been recently addressed experimentally in Ref. \cite{Ronnow}. 

Regarding our spin-boson mixture \cite{Peter}, the first specie lies in a deep optical lattice with tightly confining wells and forms a spin array; spin-up/down  corresponds to occupation by one/no atom at each site. The second specie forms a superfluid reservoir. Different species are coupled coherently via laser transitions and collisions. Whereas the laser coupling mimics a transverse field for the spins, the coupling to the reservoir sound modes induces the ferromagnetic (Ising) coupling as well as dissipation.
By measuring the critical exponents, one may confirm that the dissipative phase transition is still of second-order type where they are related by $\beta (\delta - 1) = \gamma$. The order
parameter $\langle \sigma_{iz}\rangle$ now must go continuously to zero at a second-order phase transition which is in striking contrast to the case of a single two-level system coupled to a ohmic bath. Finally, for the dissipative spin chain, the value of the dynamic critical exponent can be directly obtained from the equality $z=(\delta-1)/2 \sim 2$.

Open issues in the field include the entanglement properties as well as the spin dynamics of the dissipative quantum Ising chain. Concerning the last point, a progress in this direction has been recently
achieved in Ref. \cite{Fazio}. However, the authors have neglected the long-range spin correlations in time induced by the bath assuming that the bosons have a nonzero inverse lifetime. 

\section{One-mode superradiance model}

An ensemble of $N$ two-level atoms interacting with a radiation field has been studied
by many authors \cite{Dicke,TaynesCummings,ScullyLamb}. In the celebrated Dicke model, the atoms are assumed to be at fixed positions within a linear cavity of volume $V$ and the separations between the atoms are large enough so that the interaction among them can be ignored. However, the fact that the atoms interact with the same radiation field, they cannot be treated as independent. The importance of
 treating the radiating atoms as a {\it single} quantum system was recognized by Dicke \cite{Dicke}.
An exact solution for the Hamiltonian of $N$ identical two-level atoms interacting with a single-mode quantized radiation field at resonance was given by Tavis and Cummings \cite{TaynesCummings}. The
thermodynamic properties of the system in the limit $N$, $V\rightarrow +\infty$, $N/V\approx$ finite have been first obtained by Hepp and Lieb \cite{Hepp}. They reveal a second-order classical phase transition. The single-mode Dicke model also admits a second-order quantum phase transition \cite{Wang,Emary}. Below, we analyze this second-order quantum phase transition.

\subsection{Hamiltonian}

We study the quantum regime of the one-mode superradiance (Dicke) model \cite{Dicke} where collective and coherent behavior of the pseudospins (atoms) is induced by coupling --- with interaction $\lambda$ --- to a physically distinct {\it single-boson} subsystem. In the following, when we refer to the Dicke model we shall mean the single-mode Hamiltonian unless otherwise stated. With omission of the ${A}^2$ term for the electromagnetic field, the Hamiltonian reads \cite{Emary}:
\begin{equation}
H = \omega_s J_z + \omega a^{\dagger} a +\frac{\lambda}{\sqrt{2j}}\left(a+a^{\dagger}\right)(J_+ + J_-),
\end{equation}
where this form follows from the introduction of a collective spin operator of length $j=N/2$. The resonance condition is $\omega=\omega_s$. The thermodynamic limit of $N\rightarrow +\infty$ is thus equivalent to making the length of the pseudo-spin tend to infinity $j\rightarrow \infty$. Here, 
$\omega_s$ is the frequency splitting between the atomic levels, $\omega$ is the frequency of the field
mode, and $\kappa$ is the dipole coupling strength. In addition, the collective atomic operators satisfy angular momentum commutation relations $[J_+,J_-]=2J_z$ and $[J_{\pm},J_z]=\mp J_{\pm}$.

The Dicke model is usually considered in the quantum optics approach of the rotating-wave approximation, which is valid for small values of $\lambda$, and consists to neglect the counter-rotating
terms $a^{\dagger}J_+$ and $aJ_-$.  This makes the model integrable and simplifies the analysis.
Below, we follow Emary and Brandes, and derive exact results
without the rotating-wave approximation \cite{Emary}.
In the thermodynamic limit $(N,j)\rightarrow \infty$, the system shows a quantum phase transition at a critical coupling $\lambda_c=\sqrt{\omega \omega_s}/2$, and the system changes from a {\it normal} phase to a {\it superradiant} one. 
Superradiance means the decay of an excited population of atoms via spontaneous emission
of photons. 

In fact, it is instructive to note that the problem reduces to a two-mode problem by using the Holstein-Primakoff transformation \cite{Holstein} of the angular momentum operators $J_z=
(b^{\dagger} b -j)$, $J_+=b^{\dagger}\sqrt{2j-b^{\dagger} b}$, and $J_-=J_+^{\dagger}$; here, $b$ and $b^{\dagger}$ represent Bose operators $[b,b^{\dagger}]=1$. The Hamiltonian becomes \cite{Emary}:
\begin{equation}
H = \omega_s(b^{\dagger} b -j)  + \omega a^{\dagger} a +\lambda\left(a+a^{\dagger}\right)\left(b^{\dagger}\sqrt{1-\frac{b^{\dagger} b}{2j}} +\sqrt{1-\frac{b^{\dagger} b}{2j}}b\right).
\end{equation}

\subsection{Normal phase}

In the normal phase $(\lambda<\lambda_c)$, one can expand the square roots and this gives:
\begin{equation}
H_{n} = \omega_s b^{\dagger} b - j\omega_s +\omega a^{\dagger}a +\lambda(a+a^{\dagger})(b+b^{\dagger}).
\end{equation}
This problem of two-coupled harmonic oscillators is exactly solvable. After diagonalizing the problem, one gets two independent (effective) oscillators:
\begin{equation}
H_{n} = E_-^{(n)} c^{\dagger}_1 c_1 +E_+^{(n)} c^{\dagger}_2 c_2 +\frac{1}{2}\left(E_+^{(n)} + E_-^{(n)}-\omega-\omega_s\right) - j\omega_s.
\end{equation}
The bosonic operators $(c_1,c^{\dagger}_1,c_2,c^{\dagger}_2)$ are linear combinations of the original
operators and describe collective atom-field excitations. The energies $E_{\pm}^{(n)}$
of the two independent oscillator modes take the form \cite{Emary}:
\begin{equation}
\left(E_{\pm}^{(n)}\right)^2 = \frac{1}{2}\left(\omega^2+\omega_s^2\pm \sqrt{(\omega_s^2-\omega^2)^2+16\lambda^2\omega\omega_s}\right).
\end{equation}
It should be noted that the excitation energy $E_-^{(n)}$ is real only if $\omega^2+\omega_s^2\geq \sqrt{(\omega_s^2-\omega^2)^2+16\lambda^2\omega\omega_s}$ implying $\lambda\leq \sqrt{\omega\omega_s}/2=\lambda_c$. This underlines that $H_{n}$ is valid only for $\lambda\leq \lambda_c$, {\it i.e.}, in the normal phase. For large $j$, the ground state energy is $-j\omega_s$ whereas the excitation energies $E_{\pm}^{(n)}$ are ${\cal O}(1)$. 

In the normal phase there is a conserved quantity (parity) which is given
by: $\Pi=\exp(i\pi N_e)$ and $N_e=a^{\dagger} a + J_z +j$ represents the excitation number. It counts the total
number of excitation quanta in the system and possesses two eigenvalues $\pm 1$. After the Holstein-Primakoff transformation, the parity operator becomes $\Pi=\exp\left(i\pi[a^{\dagger} a + b^{\dagger} b]\right)$. In this formulation, there is an apparent analogy with the parity operator of a two-dimensional harmonic operator. 
The ground state has a positive parity  with an even
excitation number: this is obvious at $\lambda=0$ since the excitation number is zero. In addition, as the energy levels in this phase are non-degenerate the continuity of the ground state with increasing $\lambda$
ensures that it always has positive parity. 

\subsection{Superradiant phase}

In the superradiant phase $(\lambda>\lambda_c)$ the field and the atomic ensemble acquires macroscopic occupations, thus one has to redefine \cite{Emary}:
\begin{equation}
a^{\dagger} \rightarrow c^{\dagger}+\sqrt{m},\ b^{\dagger}\rightarrow d^{\dagger}-\sqrt{n},
\end{equation}
where $m$ and $n$ are ${\cal O}(j)$. The problem can be still diagonalized, and again one gets a theory of two decoupled oscillators:
\begin{eqnarray}
H_{s} &=& E_-^{(s)} e^{\dagger}_1 e_1 + E_+^{(s)} e^{\dagger}_2 e_2 -j\left(\frac{2\lambda^2}{\omega} +\frac{\omega_s^2 \omega}{8\lambda^2}\right) \\ \nonumber
&+&\frac{1}{2}\left(E_+^{(s)} + E_-^{(2)} -\frac{\omega_s\lambda^2}{2\lambda_c^2}\left(1+\frac{\lambda_c^2}{\lambda^2}\right)-\omega-\frac{2\lambda^2}{\omega}\left(1-\frac{\lambda_c^2}{\lambda^2}\right)\right),
\end{eqnarray} 
The expressions of the bosonic creation and annihilation operators $(e_1,e^{\dagger}_1,e_2,e^{\dagger}_2)$ can be found in the Appendix A of Ref. \cite{Emary} and the oscillator energies obey:
\begin{equation}
2\left(E_{\pm}^{(s)}\right)^2 = \frac{\omega_s^2}{\mu^2}+\omega^2\pm\sqrt{\left(\frac{\omega_s^2}{\mu^2}-\omega^2\right)^2+4\omega^2\omega_s^2},
\end{equation}
where the parameter $\mu$ is given by $\mu = \omega \omega_s/4\lambda^2 = \lambda_c^2/\lambda^2$.
Again, one can check that the excitation $E_{-}^{(s)}$ remains real as long as $\lambda>\lambda_c$, and the ground state energy is given by $-j[(2\lambda^2/\omega)+(\omega_s^2\omega/8\lambda^2)]$. It is relevant
to observe that each of every level of the total spectrum is doubly degenerate above the phase transition.
 In particular, one might also define $a^{\dagger} \rightarrow c^{\dagger}-\sqrt{m},\ b^{\dagger}\rightarrow d^{\dagger}+\sqrt{n}$, leading to the same energy spectrum. It implies that the symmetry of the ground
 state defined by the parity operator $\Pi$ is spontaneously brokem at $\lambda_c$. Nevertheless, although the global symmetry $\Pi$ is broken at the phase transition one can define a new operator $\Pi^{(s)}=\exp\left(i\pi[c^{\dagger} c + d^{\dagger} d]\right)$ which commutes with  the superradiant Hamiltonian $H_{s}$.
 
The normal phase allows a ferromagnetic ordering for the pseudospins $(\langle J_z\rangle\rightarrow -j)$, whereas in the superradiant phase, $\langle J_z\rangle$ decreases continuously.

\subsection{Second-order quantum phase transition}

The excitations of the system are given by the energies $E_{\pm}$, which describe collective modes
or polaritons in solid-state physics. As the coupling $\lambda$ approaches $\lambda_c$ one can observe
that $E_-^{(n)}=E_-^{(s)}=0$, signaling the occurrence of the quantum phase transition. In contrast, $E_+$
tends to the value $\sqrt{\omega_s^2+\omega^2}$ as $\lambda\rightarrow\lambda_c$ from either direction.
In fact, one may identify $E_-$ with the excitation energy of a photon branch and $E_+$
with the excitation energy of an atomic branch \cite{Emary}. In addition, for $\lambda\rightarrow \lambda_c$, from either direction one gets
$E_-(\lambda\rightarrow \lambda_c) \propto |\lambda_c-\lambda|^{z\nu}$,
where the dynamic critical exponent reads $z=2$ and $\nu=1/4$ is the critical exponent describing the divergence of the characteristic length $\xi= E_-^{-1/2}$. The fact that $E_-$ vanishes at $\lambda_c$ implies that the Dicke model exhibits a second-order quantum phase transition. 
The entanglement between the atoms and field diverges with the same critical exponent as the characteristic length \cite{Lambert}.
Emary and Brandes have also shown that at the quantum phase transition
the system changes from being quasi-integrable to quantum chaotic \cite{Emary}. The finite-size scaling exponents of the Dicke model have been discussed in Ref. \cite{Dusuel}. The quantum phase transition for the multi-mode (and continuum) case has been discussed in Ref. \cite{Tolkunov}.  (Tolkunov and Solenov argue that adding the $A^2$ term in the Hamiltonian, one would observe the same transition  with a corrected position of the critical point \cite{Tolkunov}.)

\subsection{Realizations}

In familiar quantum-optical systems, the frequencies $\omega$ and $\omega_s$ exceed the dipole coupling strength by many orders of magnitude. Therefore, the quantum dissipation due to atomic spontaneous emission and cavity loss is usually unavoidable, and the quantum
phase transition remains unobserved experimentally. On the other hand, the Dicke model in the quantum phase transition regime 
$(\omega_s\approx \omega \approx \lambda)$ may be realized based on the collective interaction of an ensemble of atoms with laser fields and field modes of a high-finesse optical resonator. In particular, cavity quantum electrodynamics (QED) \cite{Miller} might realize the Dicke model with parameters $\omega_s \approx \omega \approx \lambda$ that are adjustable and can in principle exceed all dissipation rates \cite{Dimer}. More precisely, Dimer {\it et al.} have proposed a well-defined  scheme based on multilevel
atoms and cavity-mediated Raman transitions to realize a Dicke model in an open system dynamics (with omission of the
$A^2$ term) \cite{Dimer}. The ensemble of atoms is simultaneously coupled to the quantized field of the optical cavity mode and the classical field of a pair of lasers. In principle, optical light from the cavity carries signatures of the critical behavior. Another scheme based on a superconducting quantum interference device coupled to a high-quality cavity supporting a single-mode photon has also been proposed in Ref. \cite{Chen}.

\section{Jaynes-Cummings lattice}

Finally, we consider a two-dimensional array of coupled optical cavities, which may be realized in circuit QED for example, each containing a single two-level atom (spin) in the photon-blockade regime. The coupling between the atom and the photons leads to an effective photon-photon {\it repulsion}; this photonic repulsion (blockade) has been shown recently by Birnbaum {\it et al.} using a single trapped atom \cite{Birnbaum}. Below, we study the resulting Jaynes-Cummings \cite{JaynesCummings} lattice-type model. Following a mean-field theory, first we show that the system
at zero temperature can undergo a characteristic Mott insulator (excitations localized on each site)
to superfluid (polaritons delocalized across the lattice) quantum phase transition \cite{Greentree}. Then, we make a rigorous comparison between the Jaynes-Cummings lattice and the Bose-Hubbard model \cite{jensKaryn}. 

\subsection{Hamiltonian}

We consider a system composed of a regular array of identical cavities. For a sufficiently large quality factor, we may restrict our treatment to a single photon mode, similar to the Dicke model above. We assume that the finite quality factor only stems from photon leakage $\kappa$ among nearest-neighbor cavities.
The Jaynes-Cummings lattice model then takes the following form \cite{Greentree}:
\begin{equation}
H=\sum_i H_i^{JC} -\kappa\sum_{\langle i;j\rangle} \left(a^{\dagger}_i a_j +a^{\dagger}_j a_i\right) - \mu\sum_i 
\left(a^{\dagger}_i a_i +\sigma_i^+ \sigma_i^-\right),
\end{equation}
where we have introduced the Jaynes-Cummings Hamiltonian\footnote{The light-atom coupling $\beta$ should not be confused with $1/k_B T$.}:
\begin{equation}
H_i^{JC} = \omega a^{\dagger}_i a_i + \omega_s \sigma_i^+ \sigma_i^- + \beta(a_i^{\dagger}\sigma_i^- +\sigma_i^+ a_i).
\end{equation}
The operator $a_i$ $(a_i^{\dagger})$ annihilates (creates) one photon in a given cavity at lattice site
$i$. Similarly $\sigma_i^{\pm}$ denote the Pauli raising and lowering operators for each two-level atom.
The bracket notation $\langle i,j\rangle$ denotes summation over nearest neighbor pairs.
The phase boundary between a Mott insulator and a superfluid phase can be determined in a grand canonical approach. To this end, a chemical potential $\mu$ is introduced. The grand canonical approach considers a situation in which particle exchange with the surrounding is permitted and is used
because of its convenience for determining the phase diagram. The boundary between Mott insulator
and superfluid phases is determined by the value of $\mu$ for which adding or removing a particle
does not require energy. In addition, here we work in the limit of the rotating-wave approximation. 

\subsection{Mott insulator-superfluid transition}

In the limit $\kappa\rightarrow 0$ (``atomic'' limit), the Hamiltonian decouples in the site index and reduces to a Jaynes-Cummings Hamiltonian $H_{\kappa=0}=H^{JC}-\mu n$ where the {\it total} excitation number on each site now obeys $n = a^{\dagger} a +\sigma^+ \sigma^-$. Since $H^{JC}$ and $n$ commute we can write
the eigen-energies of $H_{{\kappa}=0}$ as ${\cal E}_{|n\pm\rangle} = E_{|n\pm \rangle} -\mu n$, where
the usual eigen-energies of the Jaynes-Cummings Hamiltonian are given by: $E_{|n\pm\rangle}=n\omega-\Delta/2 \pm \sqrt{n\beta^2 +\Delta^2/4}$ for $n\geq 1$ and $E_{|0\rangle} = 0$ for the ground
state. Here, $\Delta=\omega-\omega_s$ is the detuning between the two-level atom and the
resonator frequency. The polariton states $|n_{\pm}\rangle = (|n,\downarrow\rangle \pm |(n-1),\uparrow\rangle)/\sqrt{2}$ are simultaneous eigenstates of the Jaynes-Cummings Hamiltonian and of the polariton number $n= a^{\dagger} a +\sigma^+ \sigma^-$. For a gien $n$, since the state $|n+\rangle$ is always higher in energy compared
to $|n-\rangle$, we can ignore it completely when focussing on ground state properties. The transition from the boundary 
$|n-\rangle$ to $|(n+1)-\rangle$ occurs when ${\cal E}_{|n -\rangle} = {\cal E}_{|(n +1)-\rangle}$
$(n=0,1,...)$. 
Therefore, this implies: $(\mu-\omega)/\beta = \sqrt{n+(\Delta/2\beta)^2} - \sqrt{n+1+(\Delta/2\beta)^2}$.
This reduces to $(\mu-\omega)/\beta=\sqrt{n}-\sqrt{n+1}$ in the resonant case $\omega=\omega_s$.
Those boundaries separate the different Mott phases for the polaritons (excitations); see Fig. 1.3 (left). These results for the atomic limit are consistent with the results of Ref. \cite{Greentree}. By increasing the hopping $\kappa$, one expects a second-order quantum phase transition from a Mott-insulating (MI) phase to a superfluid phase (SF) of polaritons which can be first described using a mean-field theory.

\begin{figure}
\resizebox{\hsize}{!}{
\includegraphics{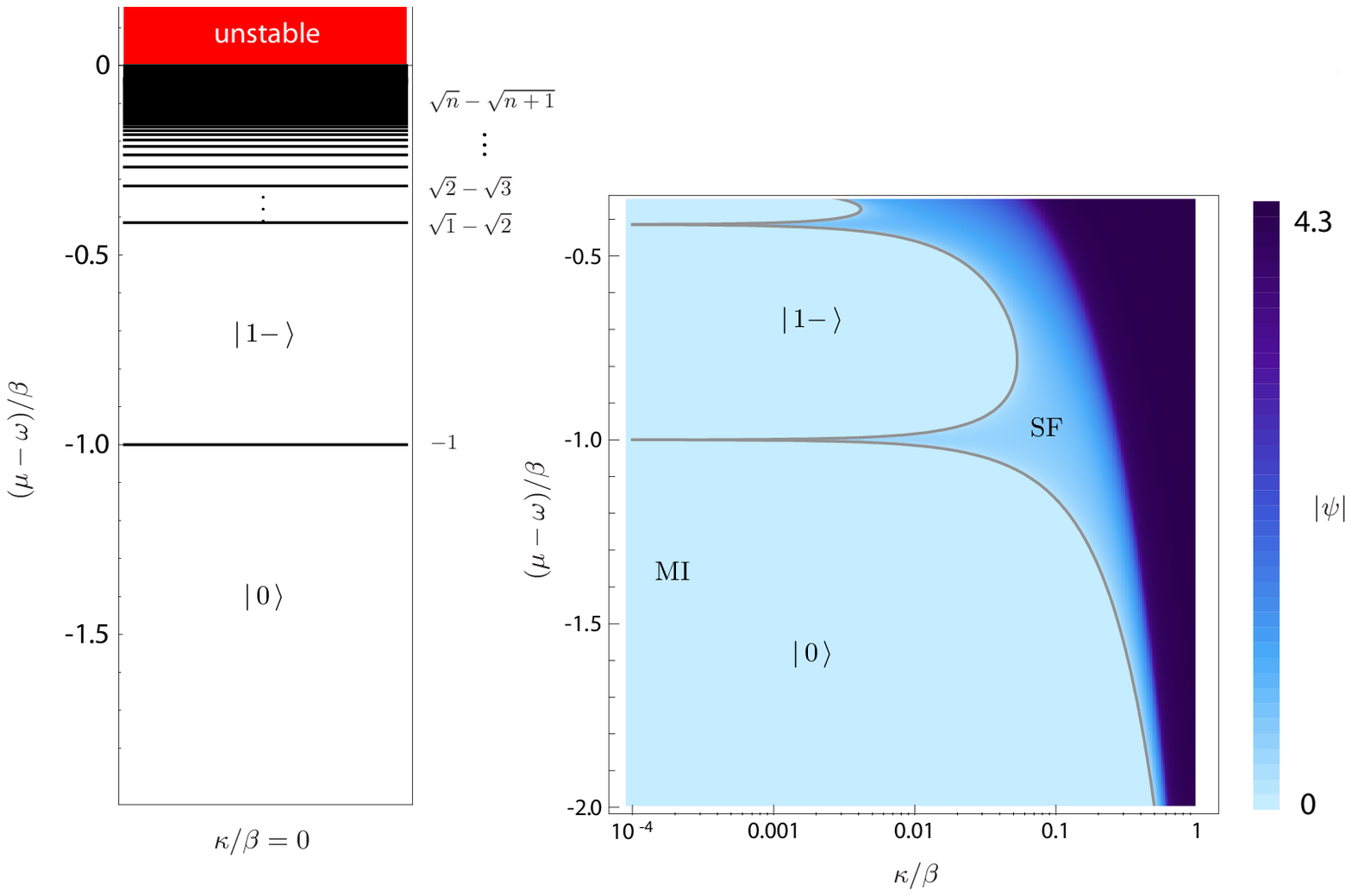}
}
\caption{(left) Ground state for $\kappa=0$ and $\Delta=0$. (right) Mean-field phase diagram.}
\label{fig:trois}
\end{figure}

Introducing the order parameter $\psi=z_c \kappa \langle a_i\rangle$ where $z_c$ is the coordination number of the lattice, the phase boundary between the MI and SF phases can be determined in way similar to to the procedure for the Bose-Hubbard model \cite{BH}. In the critical region the ground state energy can be expanded as \cite{jensKaryn}:
\begin{equation}
E_G(\psi)= E_G^{mf} + r|\psi|^2 +\frac{1}{2} u|\psi|^4 + {\cal O}(|\psi|^6).
\end{equation}
This represents the standard situation of a quadratic plus quartic potential, ubiquitous in the study of mean-field phase transitions. At the mean-field level, the phase boundary is specified by the condition $r=0$. In fact, the coefficient $r$ can be expressed as $r=R_n +(z_c\kappa)^{-1}$ where $R_n$ can be obtained from second-order perturbation theory in the photon hopping \cite{Greentree,jensKaryn} leading to Fig. 1.3 (right). 
Recently, a strong-coupling theory to the phase diagram has been developed in Ref. \cite{Blatter} which allows to include the leading correction due to quantum fluctuations for the phase boundary. The phase boundary is in agreement with Quantum Monte Carlo calculations in two dimensions \cite{Ueda}.

\subsection{Spin-1/2 mapping for the polaritons}

To make a more concrete analogy with the Bose-Hubbard model, one can build a spin-1/2 mapping
for the polaritons between two adjacent Mott lobes. More precisely, the polariton operators are defined
as \cite{Angelakis} $P_{i,n\alpha}^{\dagger} = |n\alpha\rangle_i\langle 0-|_i$,
where $\alpha=\pm$ and $i$ denotes the site index. Note that these operators do not satisfy the canonical
commutation relations of creation and annihilation operators. Now, close to a degeneracy point between two Mott-insulating lobes (polariton occupation numbers $n-1$ and $n$) in the atomic limit $\kappa/\beta\ll 1$, one can build an effective model for low-energy states in this regime from the two relevant product states $|(n-1)-\rangle^{\otimes_j}$ and $|n-\rangle^{\otimes_j}$. Within this truncated Hilbert space, the Hamiltonian $H_n^{eff}$ can be re-written as a spin-lattice $XX$ model. For this purpose, one can introduce spin-1/2 operators $\sigma_i^+ = P^{\dagger}_{i,n-} P_{i,(n-1)-}$ \cite{Angelakis}.
Within this definition, one can check that $\sigma_i^+$ and its Hermitian conjugate $\sigma_i^-$ obey standard commutation and anticommutation rules for the Pauli lowering and raising operators \cite{jensKaryn}. The 
XX model describing the physics close to the degeneracy points of the Jaynes-Cummings lattice model
is given by \cite{jensKaryn,Angelakis}:
\begin{eqnarray}
H_n^{XX} &=& \frac{1}{2}\left({\cal E}_{|n-\rangle} - {\cal E}_{|(n-1)-\rangle}\right)\sum_i \sigma_i^z \\ \nonumber
&-& \frac{1}{2}\kappa t_n^2 \sum_{\langle i;i'\rangle} \left(\sigma_i^x \sigma_{i'}^x + \sigma_i^y \sigma_{i'}^y\right).
\end{eqnarray}
The conversion amplitudes $t_n$ are given in Ref. \cite{jensKaryn}. This effective description is reminiscent of the Bose-Hubbard model between two Mott lobes.

\subsection{Field theory approach of the transition}

In fact, a field-theory formulation can also been built by analogy to the Bose-Hubbard model \cite{jensKaryn}. After performing a usual Hubbard-Stratonovich transformation to decouple the hopping term, then
one obtains the following effective Lagrangian for the auxiliary fields $\psi^*(x,\tau)$ and $\psi(x,\tau)$ (which are proportional to the order parameter $\langle a_i\rangle$ and hence are small in the critical region):
\begin{eqnarray}
{\cal L}_{eff} = K_0 + K_1 \psi^* \frac{\partial \psi}{\partial \tau} + K_2 \left|\frac{\partial \psi}{\partial \tau}\right|^2 + K_3\left| \nabla \psi\right|^2 + \tilde{r}|\psi|^2 +\frac{\tilde{u}}{2}|\psi|^4...
\end{eqnarray}
As in the case of the Bose-Hubbard model \cite{Subir,Herbut}, the coefficients $\tilde{r}$ and $\tilde{u}$ can be related to
the mean-field coefficients: $\tilde{r}=v^{-1} r$ and $\tilde{u}=v^{-1} u$ where $v$ is the volume per lattice site \cite{jensKaryn}. To obtain the coefficients $K_1$ and $K_2$ one can use the fact that the theory
must be invariant under a (time-dependent) U(1) symmetry. Using the gauge symmetry of the action 
$S_{eff} = \int_0^{\beta} d\tau \int d^d x {\cal L}_{eff}$, then one obtains the precise (exact) equalities \cite{jensKaryn}:
\begin{equation}
K_1 = \frac{\partial \tilde{r}}{\partial (\omega-\mu)}\hskip 0.5cm \hbox{and}\hskip 0.5cm K_2 = -\frac{1}{2}\frac{\partial^2 \tilde{r}}{\partial
(\omega-\mu)^2}.
\end{equation}
Therefore, whenever the coefficient $K_1$ vanishes, then the phase transition changes its universality class since the dynamical critical exponent $z=1$. The physics of these multicritical points is in complete
analogy to the corresponding physics of the Bose-Hubbard model \cite{BH,Subir,Herbut}. In contrast, for $K_1\neq 0$, we have a rather different field theory and the dynamical critical exponent associated with
the quantum phase transition is $z=2$.

On the other hand, the Jaynes-Cummings lattice has one additional parameter, the energy scale for the
atoms $\omega_s$. Thus, the phase boundary is a two-dimensional surface in the space spanned
by the parameters $\omega$, $\omega_s$, and $\kappa$, and the condition $K_1=0$ defines curves
on the phase boundary, whose position is completely determined by $\partial R_n/\partial \omega|_{\omega_s} =0$ \cite{jensKaryn}. For each Mott lobe, there is one such multicritical curve, along which the universality class changes in a similar way to the Bose-Hubbard model. It is
interesting to observe that the evidence for these multicritical curves is currently discussed controversially. Schmidt and Blatter \cite{Blatter} have presented evidence for the
presence of such multicritical points which is consistent with the results from field theory shown above \cite{jensKaryn}. However, Zhao {\it et al.} \cite{Ueda} argue for the absence of such multicritical points based on quantum Monte Carlo simulations. On the other hand, Zhao {\it et al.} have defined the dynamic critical exponent $z$ from the superfluid stiffness of the photons only (and not of the polaritons); in particular, the photon number cannot be conserved (at the lobe-tips).

\subsection{Realizations}

Several candidates have been proposed for an actual realization of the Jaynes-Cummings lattice, ranging from arrays of photonic band-gap cavities to circuit QED systems \cite{Plenio,Yamamoto}. A natural candidate system is the microwave strip line resonator for circuit QED \cite{Walraff}.
We consider that the realization in circuit QED is especially interesting \cite{jensKaryn}. Mostly, the basic
building block is well established and show the required Jaynes-Cummings physics \cite{Steve}. In particular, medium-size arrays (with a number of sites between 10 and 100) should not pose fundamental difficulties and ideas to prepare the system have been envisioned \cite{Angelakis}. First, the system would be prepared in the Mott insulating regime by using a global external microwave signal. Then, the system could cross the phase boundary by varying the detuning $\Delta$, whereas the hopping strength $\kappa$ is fixed by the fabrication parameters. The realization of this Jaynes-Cummings lattice would share many of the fascinating aspects of the ultracold atoms. Finally, a relevant question to address in the future is whether the effect of an additional external driving combined with the presence of dissipation may lead to a change of the universality class of the quantum phase transition.

\section{Conclusion}

In this chapter, we have provided a comprehensive and modern investigation of quantum phase transitions emerging in spin-boson Hamiltonians. The thermodynamic limit here can be achieved via a large number of bosonic modes resulting in a dissipative environment or/and through a large ensemble of spins (spin array). In the case of a single two-level system coupled to a bosonic (dissipative) environment, we have summarized recent developments on the computation of observables and spin dynamics for the ohmic case and we have discussed the  quantum phase transition in the sub-ohmic case by analogy to Ising spin chains with long-range interactions. Additionally, we have shown that a spin array coupled to the same dissipative bosonic bath via a longitudinal coupling gives rise to a dissipative quantum Ising model which may be engineered in cold atomic systems. On the other hand, the light-atom interaction also results in fascinating quantum phase transitions such as the zero-temperature superradiant transition taking place in the Dicke model when tuning the dipole coupling strength $\lambda$. The Jaynes-Cummings lattice system comprising
an array of optical cavities each containing a single atom also allows to realize a superfluid-Mott insulator transition for polaritons. The Dicke model and the Jaynes-Cummings lattice may be implemented in QED cavity systems. In this sense, setups of electromagnetic resonators  
could embody a novel class of quantum simulators of condensed-matter Hamiltonians.

We acknowledge C.-H. Chung, Ph. Doucet-Beaupr\' e, W. Hofstetter, A. Imambekov, J. Koch, A. Kopp, M.-R. Li, P. P. Orth, D. Roosen, I. Stanic, M. Vojta, and P. W\" olfle for fruitful collaborations. We also thank M. B\" uttiker, S. Girvin, L. Glazman, D. Goldhaber-Gordon and G. Zarand for stimulating discussions. The research on entanglement in many-body systems such as spin-boson systems is supported by NSF under the grant DMR-0803200. We also acknowledge the support from NSF through the Center for Quantum Information Physics at Yale (DMR-0653377) and from DOE under the grant DE-FG02-08ER46541.

\section*{References}
\addcontentsline{toc}{chapter}{References}

\end{document}